\documentclass{article}
\usepackage{amsmath,amssymb,amsopn,amsfonts,bbold,bm,empheq,amscd,accents,framed,delarray,epic}
\usepackage[mathscr,mathcal]{euscript}

\newcommand{\be}{\begin{equation}}
\newcommand{\ee}{\end{equation}}
\newcommand{\bmat}{\begin{pmatrix}}
\newcommand{\emat}{\end{pmatrix}}

\begin{document}

\title{On the Combination Procedure of Correlated Errors}
\author{Jens Erler\footnote{erler@fisica.unam.mx} \\
\normalsize\em Instituto de F\'isica, Universidad Nacional Aut\'onoma de M\'exico \\
\normalsize\em Apartado Postal 20--364, M\'exico D.F. 01000, M\'exico}

\maketitle
 

\begin{abstract} 
When averages of different experimental determinations of the same quantity are computed,
each with statistical and systematic error components,
then frequently the statistical and systematic components of the combined error are quoted explicitly. 
These are important pieces of information since statistical errors scale differently and often more favorably with the sample size
than most systematical or theoretical errors. 
In this communication we describe a transparent procedure 
by which the statistical and systematic error components of the combination uncertainty can be obtained. 
We develop a general method and derive a general formula for the case of Gaussian errors with or without correlations. 
The method can easily be applied to other error distributions, as well.
For the case of two measurements, we also define disparity and misalignment angles, 
and discuss their relation to the combination weight factors.
\end{abstract}

\section{Introduction}
Error propagation as well as the averaging of results of individual measurements
--- at least in the context of strictly Gaussian errors, possibly with statistical or systematic correlations ---
are straightforward, they are covered in many textbooks\footnote{For a readable and practical treatise 
from the Bayesian point of view see Ref.~\cite{Gelmanetal}.
This reference strikes a welcome balance between theoretical background and applied 
data analysis offering more than recipes.
For an upcoming textbook on data analysis in Particle Physics, see Ref.~\cite{Lista:2016tva}.}, and there seems to be no open issue,
because all that is required is multi-variate analysis applied to normal distributions.
It is the more surprising, that --- to the best of the authors knowledge ---
no explicit analytical expression is available that serves to compute for a given set of measurements 
of some quantity with individual, generally correlated, errors of statistical and systematic nature, 
the statistical and systematic components of the uncertainty of the average.

That is to say, while in the Gaussian context it is clear how to obtain the average including its uncertainty,
and that the total error ought to be the quadratic sum of the statistical and systematic 
(and perhaps other such as theoretical) error components,
formulae of these individual components and their derivation have not received much attention.
But the proper disclosure of the statistical (random) error component compared to the systematic uncertainty can be of importance.
For example, in the context of the design of future experimental facilities it is crucial to know
how much precision (say, over the world average) can be gained by simply generating larger data samples,
in contrast to possible technological or scientific breakthroughs.
The systematic error is, of course, more troublesome as it cannot be reduced\footnote{Simply collecting more data often helps 
to reduce even the systematic error component, because some error sources 
that are nominally classified as systematic may be tracable to represent themselves statistical measurements.
Moreover, with growing statistics one may restrict oneself to cleaner data by imposing stronger selection criteria (cuts). 
Nevertheless, the $N^{-1/2}$ scaling of the statistical component is usually out of reach.}
as straightforwardly by increasing the sample size $N$.

In the next Section, we review the standard procedure 
to average a number of experimental determinations of some observable quantity. 
We also mention an approximate method to obtain the statistical and systematic error of an average of  
similar experiments where the ratio of the systematic to statistical components are comparable,
or where the statistical error is dominant. 

Then we turn to the main point, the exact determination of the individual error components of an average.
We show that in the absence of correlations the square of the statistical error or any other type of uncertainty 
is weighted by the fourth power of the total errors. 

Then we turn to correlations, starting with the simplest case of two measurements 
for which we introduce the concepts of disparity and misalignment angles.
Finally, we present exact relations for the case of more than two measurements, 
and address some problems that arise when new measurements are added to an existing average iteratively.

\section{Simplified Procedures}
Suppose one is given a set of measurements of some quantity $v$, with central values $v_i$,
statistical (random) errors $r_i$ and total systematic errors $s_i$.
For simplicity, we are going to assume that the $r_i$ and $s_i$ are Gaussian distributed 
(the generalization to other error distributions is straightforward)
in which case the total errors of the individual measurements are given by
\be
\label{ti}
t_i = \sqrt{r_i^2 + s_i^2}\ .
\ee
If we furthermore temporarily assume that the measurements are uncorrelated, 
then the central value $\bar v$ of their combination is given by the precision weighted average,
\be
\label{vbar}
\bar v = {\sum_i v_i t_i^{-2} \over \sum_i t_i^{-2}} \ ,
\ee
with total error
\be
\label{tbar}
\bar t = {1 \over \sqrt{\sum_i t_i^{-2}}} \ .
\ee
Similarly, the statistical component $\bar r$ of $\bar t$ can often be approximated by
\be
\label{rbar}
\bar r = {1 \over \sqrt{\sum_i r_i^{-2}}} \ .
\ee
The systematic component $\bar s$ of $\bar t$ is then obtained from
\be
\label{sbar}
\bar s = \sqrt{\bar t^2 - \bar r^2} \ .
\ee
For example, two measurements with
\be
\label{example}
r_1 = s_2 = 30, \qquad\qquad\qquad
r_2 = s_1 = 40, \qquad\qquad\qquad
t_1 = t_2 = 50,
\ee
would result in
\be
\label{exampleresult}
\bar r = 24, \qquad\qquad\qquad
\bar s = \sqrt{674} \approx 26 \approx \bar r.
\ee
Notice, that while the individual errors in Eq.~(\ref{example}) are symmetric under the simultaneous exchange
of the statistical and systematic errors (we recall that all $r_i$ and $s_i$ are assumed Gaussian)
and the labels of the two measurements, 
the result (\ref{exampleresult}) does not exhibit the corresponding symmetry which would imply $\bar r = \bar s$ exactly.
The exact procedure introduced in the next section would indeed yield $\bar r = \bar s$ in this example.

Now consider the case where one of the systematic errors, say $s_1$, is larger and eventually $s_1 \to \infty$.
Then the weight of the first measurement approaches zero, and $\bar t \to t_2$, as expected.
However, one would also expect that $\bar r \to r_2$ and $\bar s \to s_2$, 
while instead $\bar r < r_2$ remains constant and $\bar s \to \sqrt{1924} \approx 44 > s_2$.
Thus, one would face the unreasonable result that averaging some measurement with an irrelevant constraint (with infinite uncertainty)
will decrease (increase) the statistical (systematic) error component, leaving only the total error invariant.
In other words, if in a set of measurements there is one with negligible statistical error, 
then the average would also have vanishing statistical error, 
regardless of how unimportant that one measurement is compared to the others.
Clearly, Eq.~(\ref{rbar}) is then unsuitable even as an approximation.

\begin{table}[t]
\begin{center}
\begin{tabular}{|c|c|c|c|c|c|}
\hline
experiment & $v_i$ & $r_i$ & $s_i$ & $t_i$ & $u_i$ \\
\hline
ALEPH  & 0.1451 & 0.0052 & 0.0029 & 0.0060 & 0.0057 \\
DELPHI & 0.1359 & 0.0079 & 0.0055 & 0.0096 & 0.0095 \\
L3          & 0.1476 & 0.0088 & 0.0062 & 0.0108 & 0.0106 \\
OPAL     & 0.1456 & 0.0076 & 0.0057 & 0.0095 & 0.0094 \\
\hline
combination & 0.1439 & 0.0035 & 0.0026 & 0.0043 & 0.0040 \\
\hline
\end{tabular}
\caption{\label{tab}
Central value, $v_i$, random error, $r_i$, systematic uncertainty, $s_i$, total error, $t_i$, 
and uncorrelated error component, $u_i$, for measurements of the quantity ${\cal A}_\tau$ by each
of the four LEP collaborations, as well as the corresponding numbers for the combined result.}
\end{center}
\end{table}

One can easily extend these consideration to the case where the individual measurements
have a {\em common\/} contribution $c$ entering the systematic error.
The precision weighted average (\ref{vbar}) and total error (\ref{tbar}) are then to be replaced by
\be
\label{vtbar}
\bar v = {\sum_i v_i u_i^{-2} \over \sum_i u_i^{-2}}\ ,  \qquad\qquad\qquad
\bar t = \sqrt{{1 \over \sum_i u_i^{-2}} + c^2},
\ee
where the uncorrelated error components are given by
\be
\label{ubar}
u_i = \sqrt{t_i^2 - c^2}, \qquad\qquad\qquad
\bar u = \sqrt{\bar t^2 - c^2} = {1\over \sqrt{\sum_i u_i^{-2}}}\ ,
\ee
and where $t_i^2 \geq 0$ requires $c^2 \geq - u_i^2$ for all $i$.

The general case of correlated errors will be dealt with later,
but we note that the case of two measurements with Pearson's correlation coefficient $\rho$
can always be brought to this form with $c^2$ given by
\be
\label{2dim}
c^2 = t_i^2  - u_i^2 = \rho\, t_1 t_2 = \rho\, \sqrt{u_1^2 + c^2} \sqrt{u_2^2 + c^2}.
\ee
A proper (normalizable) probability distribution requires $|\rho| \leq 1$,
so that from Eq.~(\ref{2dim}),
\be
c^2 \geq - {u_1^2 u_2^2 \over {u_1^2} + {u_2^2}}\ ,
\ee
guaranteeing that $\bar t$ is real.
On the other hand, $u_1$ or $u_2$, as well as $\bar u$, in Eq.~(\ref{ubar}) may become imaginary provided that
\be\label{negativeweight}
\rho > {t_1\over t_2}\qquad\qquad\qquad 
{\rm or} \qquad\qquad\qquad
\rho > {t_2\over t_1}
\ee
in which case the first or second measurement, respectively, contributes with negative weight,
and $\bar v$ lays no longer between $v_1$ and $v_2$.
In this situation, one rather (but equivalently) regards the measurement with a negative weight 
as a measurement of some nuisance parameter related to $c$.
Replacing the inequalities~(\ref{negativeweight}) by equalities, gives rise to an infinite weight 
(one of the $u_i = 0$) as well as $\bar u = 0$ and $\bar t = c$.

As a concrete example, each of the four experimental collaborations at the $Z$ boson factory 
LEP~1~\cite{Heister:2001uh,Abreu:1999wv,Acciarri:1998vg,Abbiendi:2001km} has measured some quantity ${\cal A}_\tau$ 
(related to the polarization of final-state $\tau$ leptons produced in $Z$ decays)
with the results shown in the Table below.
A number of uncertainties affected the four measurements in a similar way, 
leading to a relatively weak correlation matrix~\cite{ALEPH:2005ab} which, 
while not quite corresponding to the form (\ref{vtbar}), (\ref{ubar}),
can be well approximated by it when using the average 
of the square root of the off-diagonal entries of the covariance matrix $c \approx 0.0016$.

The values in the last line are $\bar v$, $\bar r$, $\bar s$, $\bar t$ and $\bar u$ as calculated
from Eqs.~(\ref{rbar}), (\ref{sbar}), (\ref{vtbar}) and (\ref{ubar}).
$\bar v$, $\bar r$ and $\bar s$ agree with Table~4.3 and $\bar t$ agrees with Eq.~(4.9) of 
the LEP combination in Ref.~\cite{ALEPH:2005ab}.

Table~\ref{tab2} shows the more recent example of the determination of the weak mixing angle~\cite{Aad:2015uau}
which is based on purely central (CC) electron events, events with a forward electron (CF), as well as muon pairs.
Here the average of the off-diagonal entries of the covariance matrix amounts to $c \approx 0.0010$.
This is an example where the dominant uncertainty is from common systematics,
namely from the imperfectly known parton distribution functions affecting the three channels
in very similar ways. 

We will return to these examples after deriving exact alternatives to formula~(\ref{rbar}).

\begin{table}[t]
\begin{center}
\begin{tabular}{|c|c|c|c|c|c|}
\hline
channel & $v_i$ & $r_i$ & $s_i$ & $t_i$ & $u_i$ \\
\hline
CC electron & 0.2302 & 0.0009 & 0.0013 & 0.0016 & 0.0012 \\
CF electron & 0.2312 & 0.0007 & 0.0013 & 0.0015 & 0.0011 \\
\hline
all electron  & 0.2308 & 0.0006 & 0.0012 & 0.0013 & 0.0008 \\
muon          & 0.2307 & 0.0009 & 0.0012 & 0.0015 & 0.0011 \\
\hline
all lepton    & 0.2307 & 0.0005 & 0.0011 & 0.0012 & 0.0007 \\
\hline
\end{tabular}
\caption{\label{tab2}
Same as Table~\ref{tab}, but for the weak mixing angle determinations by ATLAS.}
\end{center}
\end{table}

\section{Derivatively Weighted Errors}
Our starting point is the basic property of a statistical error to scale as $N^{-1/2}$ with the sample size.
To implement this, we rewrite Eq.~(\ref{ti}) as
\be
\label{tieps}
t_i = \sqrt{\epsilon^2 r_i^2 + s_i^2} \Big{|}_{\epsilon=1}\ .
\ee
Thus, the statistical error satisfies the relation,
\be
\label{rieps}
r_i = \sqrt{t_i {d t_i\over d\epsilon}\Big{|}_{\epsilon=1}}\ .
\ee
In the absence of correlations we can use Eq.~(\ref{tbar}),
and demand that analogously,
\be
\label{rbareps}
\bar r^2 = \bar t\, {d \bar t \over d\epsilon} \Big{|}_{\epsilon=1} =
\bar t^{\, 4} \sum_i t_i^{-3} {d t_i\over d\epsilon} \Big{|}_{\epsilon=1} =
\sum_i r_i^2 \left( {\bar t \over t_i} \right)^4.
\ee
Notice, that Eq.~(\ref{rbar}) can be recovered from Eq.~(\ref{rbareps}) upon substituting $t_i \to r_i$ and $\bar t \to \bar r$.
Eq.~(\ref{rbareps}) means that the {\em relative\/} statistical error of the combination, $\bar x$,
is given by the precision weighted average
\be
\bar x^2 = {\sum_i x_i^2 t_i^{-2} \over \sum_i t_i^{-2}} \ ,
\ee
where
\be
x_i \equiv {r_i \over t_i}\ , \qquad\qquad\qquad
\bar x \equiv {\bar r \over \bar t}\ .
\ee
Furthermore, giving the systematic components a similar treatment, we find
\be
\label{sbareps}
\bar s^2 = \sum_i s_i^2 \left( {\bar t \over t_i} \right)^4,
\ee
so that the expected symmetry between the two types of uncertainty becomes manifest,
and moreover, Eq.~(\ref{sbar}) now follows directly from Eqs.~(\ref{rbareps}) and (\ref{sbareps}),
rather than being enforced.
The central result is that for uncorrelated errors, the squares of the statistical and systematic 
components (or those of any other type) of an average are the corresponding individual squares 
weighted by the inverse of the fourth power of the individual total errors,
or equivalently, weighted by the square of the individual precisions $t_i^{-2}$.

Returning to the case where the only source of correlation is a common contribution $c \neq 0$
equally affecting all measurements, we find from Eq.~(\ref{vtbar}),
\be\label{rbar2}
\bar r^2 = \sum_i r_i^2 \left( {\bar u \over u_i} \right)^4, \qquad\qquad\qquad
\bar y^2 = {\sum_i y_i^2 u_i^{-2} \over \sum_i u_i^{-2}}\ ,
\ee
where
\be
y_i \equiv {r_i \over u_i}\ , \qquad\qquad\qquad
\bar y \equiv {\bar r \over \bar u} \ .
\ee
Applied to the case of ${\cal A}_\tau$ measurements we now find
\be
\bar r = 0.0035,  \qquad\qquad\qquad
\bar s = \sqrt{\bar t^2 - \bar r^2} = 0.0026,
\ee
which agree not exactly, but within round-off precision with the approximate numbers in Table~\ref{tab}.

\section{Bivariate Error Distributions}
As a preparation for the most general case of $N$ measurements with arbitrary correlation coefficients, 
we first discuss in some detail the case $N = 2$.
Recall that the covariance matrix in this case reads
\be
T \equiv \bmat t_1^2 & \rho t_1 t_2 \\ \rho t_1 t_2 & t_2^2 \emat \equiv
\bmat t_1^2 & c^2 \\ c^2 & t_2^2 \emat .
\ee
The precision weighted average is given by the expression,
\be\label{vbar2}
\bar v = \frac{v_1 t_2^2 + v_2 t_1^2 - (v_1 + v_2) c^2}{t_1^2 + t_2^2 - 2 c^2}
= \frac{v_1 + \omega\, v_2}{1 + \omega}\ ,
\ee
where
\be
\omega \equiv \frac{t_1^2 - c^2}{t_2^2 - c^2}\ , \qquad\qquad\qquad
c^2 = \frac{t_1^2 - \omega t_2^2}{1 - \omega}\ ,
\ee
obtained by minimizing the likelihood following a bivariate Gaussian distribution,
\be
{\cal L} \propto e^{- \chi^2/2},
\ee
where 
\be
\chi^2 = \vec v^{\, T} T^{-1} \vec v\ , \qquad\qquad\qquad
\vec v = \bmat v_1 - \bar v \\  v_2 - \bar v \emat .
\ee
The one standard deviation total error $\bar t$ is defined by
\be
\Delta\chi^2 \equiv \chi^2(\bar v + \bar t) - \chi^2(\bar v) \overset{!}{=} 1,
\ee
which results in
\be\label{tbar2}
\bar t = \sqrt{\frac{1 - \rho^2}{t_1^{-2} + t_2^{-2} - 2 \rho\, t_1^{-1} t_2^{-1}}} 
= \sqrt{\frac{t_1^2 t_2^2 - c^4}{t_1^2 + t_2^2 - 2 c^2}} 
= \sqrt{\frac{t_1^2 - \omega^2 t_2^2}{1 - \omega^2}}\ , 
\ee
or conversely, 
\be\label{c2}
\omega = \sqrt{\frac{t_1^2 - \bar t^2}{t_2^2 - \bar t^2}}\ , \qquad\qquad\qquad
c^2 = \bar t^2 - \sqrt{(t_1^2 - \bar t^2)(t_2^2 - \bar t^2)}\ .
\ee
Eq.÷(\ref{c2}) is useful in practice if one needs to recover the correlation between 
a pair of measurement uncertainties and their combination error.

We now turn to the generalization of Eq.~(\ref{rbareps}) in the presence of a systematic correlation.
When applying our method of derivatively weighted errors to Eq.~(\ref{tbar2}) it is important to keep 
$c^2 = \rho t_1 t_2$ fixed (this would be different in the presence of a statistical correlation). 
Doing this, we obtain
\be\label{rbar3}
\bar r = \frac{\sqrt{r_1^2 (t_2^2 - c^2)^2 + r_2^2 (t_1^2 - c^2)^2}}{t_1^2 + t_2^2 - 2 c^2} =
\frac{\sqrt{r_1^2 + \omega^2 r_2^2}}{1 + \omega}\ .
\ee
For the systematic component we find
\be
\bar s = \frac{\sqrt{s_1^2 + 2 \omega c^2 + \omega^2 s_2^2}}{1 + \omega}\ ,
\ee
and we also note that
\be
\bar u ^2 = \frac{\omega}{1 - \omega^2} (t_2^2 - t_1^2).
\ee

More generally, one can compute the error contribution $\bar q$ of any individual source of uncertainty $q$ to the total error as
\be\label{qbar}
\bar q = \frac{\sqrt{q_1^2 + 2 \omega c_q^2 + \omega^2 q_2^2}}{1 + \omega}\ ,
\ee
where $c_q^2$ is the contribution of $q$ to $c^2$ with the constraint
\be
\sum_q c_q^2 = c^2.
\ee
If the two uncertainties $q_i$ are fully correlated or anti-correlated between the two measurements,
then 
\be
c_q^2 = \pm q_1 q_2\ ,  \qquad\qquad\qquad
\bar q = \frac{q_1 \pm \omega q_2}{1 + \omega}\ ,
\ee
where the minus sign corresponds to anti-correlation.

The formalism is now general enough to allow statistical correlations, as well.
As we will illustrate later, knowing all the $\bar q$ is particularly useful 
if one wishes to successively include additional measurements to a combination --- one by one --- 
rather than having to deal with a multi-dimensional covariance matrix.  
This situation frequently arises in historical contexts when new measurements
add information to a set of older ones, rather than superseding them.
But there is a problematic issue with this, which apparently is not widely appreciated.

\section{Disparity and Misalignment Angles}
Continuing with the case of two measurements, we can relate $\rho$ to the rotation angle necessary to diagonalize the matrix $T$. 
If we define an angle $\beta$ quantifying the {\em disparity\/} of the total errors of two measurements through
\be
\tan\frac{\beta}{2} \equiv \frac{t_1 - t_2}{t_1 + t_2}\ ,
\ee
then the diagonal from of $T$ is $R T R^T$ with
\be
R \equiv \bmat \cos\frac{\alpha}{2} & \sin\frac{\alpha}{2} \\ - \sin\frac{\alpha}{2} & \phantom{-} \cos\frac{\alpha}{2} \emat 
\ee
and
\be\label{diagangles}
\tan\alpha = \rho \cot\beta,
\ee
where
\be
- \frac{\pi}{2} \leq \alpha \leq\frac{\pi}{2}\ , \qquad\qquad\qquad
- \frac{\pi}{2} \leq \beta \leq\frac{\pi}{2}\ .
\ee

The angle $\alpha$ may be interpreted as a measure of the {\em misalignment\/} of the two measurements 
with respect to the primary observable of interest $v$.
Uncorrelated measurements of $v$ are aligned ($\rho = \alpha = 0$),
while the case $|\rho| \gg |\tan\beta|$ is reflective of a high degree of misalignment.
Indeed, in the extreme case where $\beta = 0$ ($|\alpha| = 90^\circ$)
two correlated measurements ($\rho \neq 0$) of the same quantity $v$
are equivalent to two uncorrelated measurements, only one of which having any sensitivity to $v$ at all.
To reach the decorrelated configuration involves subtle cancellations between correlations and anti-correlations
of the statistical and systematic error components of the original measurements.  

We can now express the weight factor $\omega$ in terms of the disparity and misalignment angles $\beta$ and $\alpha$,
\be
\omega = \frac{1 + \sin\beta (1 - \tan\alpha)}{1 - \sin\beta (1 + \tan\alpha)} =
\frac{1 + \sin\beta - \rho \cos\beta}{1 - \sin\beta - \rho \cos\beta} \ .
\ee
In the case $\rho = \alpha = 0$ this reduces to
\be\label{omega}
\omega = \tan^2 \left( \frac{\beta}{2} + \frac{\pi}{4} \right),
\ee
and Eq.~(\ref{vbar2}) now reads
\be\label{vbar3}
\bar v = \frac{v_1 + v_2}{2} - \sin\beta\, \frac{v_1 - v_2}{2}\ .
\ee
One can write equations of the form~(\ref{omega}) and~(\ref{vbar3}) for $\rho \neq 0$, as well,
with shifted angles $\bar\beta$ related to $\beta$ by
\be
\csc\bar\beta = \csc\beta - \tan\alpha\ .
\ee
However, this ceases to work out in the presence of a negative weight ($\omega < 0$),
in which case one would need to replace the trigonometric by the hyperbolic functions.

\section{Multivariate Error Distributions}
To treat cases of more than two measurements with generic correlations,
one can choose one of two strategies. 
Either one effectively reduces the procedure to cases of just two measurements 
(in general at the price of some precision loss)
by iteratively including additional measurements,  
or one deals with a multi-dimensional covariance matrix. 

We first discuss the latter approach, starting with the trivariate case where
\be\label{cov3}
T \equiv 
\bmat t_1^2 & \rho_3 t_1 t_2 & \rho_2 t_1 t_3 \\ \rho_3 t_1 t_2 & t_2^2 & \rho_1 t_2 t_3 \\ \rho_2 t_1 t_3 & \rho_1 t_2 t_3 & t_3^2 \emat
\equiv \bmat t_1^2 & c^2_3 & c^2_2 \\ c^2_3 & t_2^2 & c^2_1 \\ c^2_2 & c^2_1 & t_3^2 \emat
\ee
The average can be written as
\be
\bar v = \frac{\omega_1 v_1 + \omega_2 v_2 + \omega_3 v_3}{\omega_1 + \omega_2 + \omega_3}
\ee
with
\be
\omega_1 \equiv (t_2^2 - c_3^2)(t_3^2 - c_2^2) - (c_1^2 - c_2^2)(c_1^2 - c_3^2)\ ,
\ee
\be
\omega_2 \equiv (t_1^2 - c_3^2)(t_3^2 - c_1^2) - (c_2^2 - c_1^2)(c_2^2 - c_3^2)\ ,
\ee
\be
\omega_3 \equiv (t_1^2 - c_2^2)(t_2^2 - c_1^2) - (c_3^2 - c_1^2)(c_3^2 - c_2^2)\ .
\ee
The total error is given by
\be\label{tbareps3}
\bar t = \sqrt{\frac{\det T}{\omega_1 + \omega_2 + \omega_3}}\ ,
\ee
and for its statistical and systematic components we find (in the absence of statistical correlations),
\be\label{rbareps3}
\bar r = \frac{\sqrt{\omega_1^2 r_1^2 + \omega_2^2 r_2^2 + \omega_3^2 r_3^2}}{\omega_1 + \omega_2 + \omega_3}\ , 
\qquad\qquad\qquad
\bar s = \frac{\sqrt{\sum_i \omega_i^2 s_i^2 + \sum_{i \neq j} \omega_i \omega_j T_{ij}}}{\omega_1 + \omega_2 + \omega_3}\ ,
\ee
respectively.
The generalization of Eq.~(\ref{qbar}) is now also straightforward.
{\em E.g.}, in the case of 100\% correlation between the three measurements we have,
\be\label{qbar3}
\bar q = \frac{\omega_1 q_1 + \omega_2 q_2 + \omega_3 q_3}{\omega_1 + \omega_2 + \omega_3}\ .
\ee

Analogous expressions hold for cases with $N > 3$ measurements. 
For example, the covariance matrix for the case of $N=4$ reads
\be
T \equiv \bmat t_1^2 & c_{12}^2 & c_{13}^2 & c_{14}^2 \\
c_{12}^2 & t_2^2 & c_{23}^2 & c_{24}^2 \\
c_{13}^2 & c_{23}^2 & t_3^2 & c_{34}^2 \\
c_{14}^2 & c_{24}^2 & c_{34}^2 & t_4^2 \emat .
\ee
All that remains to be computed are the weight factors $\omega_i$.
We found a convenient expression for them, {\em e.g.},
\be
\omega_1 = \begin{vmatrix} t_2^2 - c_{12}^2 & c_{23}^2 - c_{12}^2 & c_{24}^2 - c_{12}^2 \\ 
c_{23}^2 - c_{13}^2 & t_3^2 - c_{13}^2 & c_{34}^2 - c_{13}^2 \\ 
c_{24}^2 - c_{14}^2 & c_{34}^2 - c_{14}^2 & t_4^2 - c_{14}^2 \end{vmatrix} .
\ee
Thus, the $\omega_i$ can be obtained by computing the determinant of a matrix
which is constructed by subtracting the $i^{th}$ column from each of the other columns 
(or the $i^{th}$ row from each of the other rows) 
and then removing the $i^{th}$ row and column.
The reader is now equipped to handle cases of any $N$ exactly.

The alternative strategy to compute averages is to add more measurements iteratively. 
We illustrate this using the example of the ATLAS results on the weak mixing angle (see Table~\ref{tab2}).
Besides the statistical error (there were no statistical correlations) there were seven sources
of systematics, six of which correlated between at least two channels. 
The breakdown of these uncertainties as quoted by the ATLAS Collaboration~\cite{Aad:2015uau} are shown in Table~\ref{tab3}.

\begin{table}[t]
\begin{tabular}{|c|c|c|c|c|r|r|c||c|}
\hline
& CC $e^-$ & CF $e^-$ & all $e^-$ & $\mu^-$ & Eq.~(\ref{qbar}) & Eq.~(\ref{qbar3}) & Ref.~\cite{Aad:2015uau} & $\Delta c_q^2$ \\
\hline 
central value & 0.2302 & 0.2312 & 0.23076 & 0.2307 & 0.23074 & 0.23075 & 0.2308 & +46.490 \\
\hline
\hline
statistics & 9 & 7 & 5.6 & 9 & 4.803 & 4.795 & 5 & $-0.156$ \\
MC statistics & 5 & 2 & 2.5 & 5 & 2.376 & 2.357 & 2 & $-0.194$ \\
$E_e$ & 4 & 6 & 3.8 & --- & 2.461 & 2.490 & 3 & $+0.308$ \\
$\Delta E_e$ & 4 & 5 & 3.3 & --- & 2.144 & 2.162 & 2 & $+0.164$ \\
$E_\mu$ & --- & --- & --- & 5 & 1.763 & 1.764 & 2 & $+0.010$ \\
\hline
PDF & 10 & 10 & 10 & 9 & 9.647 & 9.647 & 9 & $-0.011$ \\
higher orders & 3 & 1 & 1.9 & 3 & 2.272 & 2.255 & 2 & $-0.161$ \\
other sources & 1 & 1 & 1 & 2 & 1.353 & 1.353 & 2 & $+0.001$ \\
\hline
total & 15.7 & 14.7 & 12.9 & 15.0 & 11.939 & 11.938 & 11.6 & $-0.038$ \\
\hline
\end{tabular}
\caption{\label{tab3}
Central values and breakdown of uncertainties ($\times 10^4$) of the weak mixing angle determinations by ATLAS.
$E_e$ and $E_\mu$ refer to the $e^\pm$ and $\mu^\pm$ energy scales, respectively, 
while $\Delta E_e$ denotes the electron energy resolution.
The last three uncertainties from PDFs, missing higher order corrections, and other sources are taken as fully correlated,
whereas the other uncertainties are assumed uncorrelated.
The fourth column is the average of the two electron channels displayed in the second and third columns.
The $6^{th}$ column adds the muon channel ($5^{th}$ column) to them. 
The $7^{th}$ column shows the exact combination result of the three channels.
The $8^{th}$ column are the corresponding numbers quoted by ATLAS.
The interpretation of the last column is explained in the text.}
\end{table}

Good agreement is observed with Ref.~\cite{Aad:2015uau},
where the small differences are consistent with precision loss due to rounding. 
Indeed, the fact that there {\em are\/} round-off issues can already be seen from Ref.~\cite{Aad:2015uau}, 
where the quoted total error of the CF electron channel
is smaller than the sum (in quadrature) of the statistical, PDF, and other systematic errors.
A similar issue can be observed regarding the quoted combined systematic error 
which is larger than the sum in quadrature of its components.

However, there are small differences between the results from the exact procedure using Eqs.~(\ref{rbareps3}) and (\ref{qbar3})
and the iterative strategy using Eq.~(\ref{rbar3}) and (\ref{qbar}).
The reason can be traced to the asymmetric way in which the error due to higher orders enters the two electron channels. 
This induces subtle dependences of all sources of uncertainties (even those that were initially uncorrelated) on the correlated ones.  
It even affects the uncertainty induced by the finite muon energy resolution,
which does not enter the electron channels at all.
To account for this one can introduce {\em additional\/} contributions $\Delta c_q^2$ to the off-diagonal entry 
of the bivariate covariance matrix of the all-electron result and the muon channel.
These $\Delta c_q^2$ can be chosen to enforce the exact result, but it is impossible to compute them beforehand. 
In fact, they depend on the new measurement to be added (here the muon channel) 
and not just the initial measurements (here the two electron channels).
Moreover, the $\Delta c_q^2$ necessary to enforce the correct average central value $\bar v$ {\em differs\/} strongly 
from the $\Delta c_q^2$ necessary to enforce the total error $\bar t$.
This observation is a reflection of the fact that the {\em combination principle\/} can be violated~\cite{Lyons:1989gh},
which we state as the requirement that the combination of a number of measurements
must not depend on the order in which they added to the average.

Thus, the iterative procedure generally suffers from a loss of precision.
In this example the procedure gives nevertheless an excellent approximation
because the uncertainty from higher order corrections (the origin of the asymmetric uncertainty) is itself very small. 
But there are cases in which the iterative procedure does not provide even a crude approximation
and where one should use --- if possible --- the exact method based on the full covariance matrix.
Unfortunately, its construction is not always possible, {\em e.g.\/}, due to incomplete documentation of past results.
Recent discussions of related aspects of this conundrum can be found in Refs.~\cite{Nisius:2014wua,Lista:2014qia}.

\section{Summary and Conclusions}
\label{conclusions}
In summary, we have introduced a formalism (derivatively weighted errors) to derive formulas 
for random errors or any error type of uncorrelated Gaussian nature. 
We introduced what we call the disparity and misalignment angles to describe the case of two measurements,
and showed their relation to the statistical weight factors.
For the case of more than two measurements with known covariance matrix, 
we derived some explicit formulas in a form which (as far as we are aware) did not appear before.

It is remarkable, that even in the context of purely Gaussian errors and perfectly known correlations
there are intractable problems at the most fundamental statistical level. 
Specifically, they may arise even when a number of observations of the same quantity is combined and the error sources 
are recorded and the assumptions regarding their correlations are spelled out carefully.
In statistical terms, one can conclude that such a combination --- despite of all its recorded details ---
represents an {\em insufficient statistics\/} of the available information.
The inclusion of further observations of the same quantity is then in general ambiguous. 

On the other hand, there is no ambiguity in the absence of correlations 
or when any correlation is common to the set of observations to be combined.
The fact that the ambiguities disappear in certain limits then reopens the possibility of useful approximations. 
For example, if an iterative procedure has to be chosen, one should first combine measurements where the dominant correlation
is given approximately by a common contribution. 
Similarly, the measurements with small or no correlation with the other ones, are ideally kept for last.

\section{Acknowledgements}
This work received support from PAPIIT (DGAPA--UNAM) project IN106913 and from CONACyT (M\'exico) project 151234. 


\vspace{12pt}


\begin{thebibliography}{99}
\bibitem{Gelmanetal} 
A.~Gelman, J.~B.~Carlin, H.~S.~Stern and D.~B.~Rubin,
{\sl Bayesian Data Analysis} (Chapman \& Hall 1995).

\bibitem{Lista:2016tva} 
L.~Lista,
{\sl Statistical Methods for Data Analysis in Particle Physics},
{\em Lect.\ Notes Phys.}  {\bf 909}, (2016).
  
\bibitem{Heister:2001uh} 
ALEPH Collaboration: A.~Heister {\it et al.},
{\em Eur.\ Phys.\ J.} {\bf C20}, 401 (2001),
hep-ex/0104038.

\bibitem{Abreu:1999wv} 
DELPHI Collaboration: P.~Abreu {\it et al.},
{\em Eur.\ Phys.\ J.} {\bf C14}, 585 (2000).
  
\bibitem{Acciarri:1998vg} 
L3 Collaboration: M.~Acciarri {\it et al.},
{\em Phys.\ Lett.} {\bf B429}, 387 (1998).
  
\bibitem{Abbiendi:2001km} 
OPAL Collaboration: G.~Abbiendi {\it et al.},
{\em Eur.\ Phys.\ J.} {\bf C21}, 1 (2001),
hep-ex/0103045.
  
\bibitem{ALEPH:2005ab} 
ALEPH, DELPHI, L3, OPAL and SLD Collaborations, 
LEP Electroweak Working Group and SLD Electroweak and Heavy Flavour Groups: S.~Schael {\it et al.},
{\em Phys.\ Rept.}  {\bf 427}, 257 (2006),
hep-ex/0509008.

\bibitem{Aad:2015uau} 
ATLAS Collaboration: G.~Aad {\it et al.}, 
arXiv:1503.03709 [hep-ex].

\bibitem{Lyons:1988rp} 
L.~Lyons, D.~Gibaut and P.~Clifford,
{\em Nucl.\ Instrum.\ Meth.} {\bf A270}, 110 (1988).

\bibitem{Valassi:2003mu} 
A.~Valassi,
{\em Nucl.\ Instrum.\ Meth.} {\bf A500}, 391 (2003).

\bibitem{Lyons:1989gh} 
L.~Lyons, A.~J.~Martin and D.~H.~Saxon,
{\em Phys.\ Rev.} {\bf D41}, 982 (1990).

\bibitem{Nisius:2014wua} 
R.~Nisius,
{\em Eur.\ Phys.\ J.} {\bf C74}, 8, 3004 (2014),
arXiv:1402.4016 [physics.data-an].

\bibitem{Lista:2014qia} 
L.~Lista,
{\em Nucl.\ Instrum.\ Meth.} {\bf A764}, 82 (2014), 
arXiv:1405.3425 [physics.data-an].

\end{thebibliography}
\end{document}